\documentclass[12pt,a4paper]{article}

\usepackage{amsmath}
\usepackage{amssymb}
\usepackage{epsfig}

\newcommand{\dgr}{^{\rm o}}

\def\nostrocostrutto#1\over#2{\mathrel{\mathop{\kern 0pt \rlap 
  {\raise.2ex\hbox{$#1$}}}
  \lower.9ex\hbox{\kern-.190em $#2$}}}
\newcommand{\qqbar}{$q\overline q\ $}
\newcommand{\degr}{{\rm o}}
\catcode`@=11
\newcount\@tempcntc
\def\@citex[#1]#2{\if@filesw\immediate\write\@auxout{\string\citation{#2}}\fi
  \@tempcnta\z@\@tempcntb\m@ne\def\@citea{}\@cite{\@for\@citeb:=#2\do
    {\@ifundefined
       {b@\@citeb}{\@citeo\@tempcntb\m@ne\@citea\def\@citea{,}{\bf ?}\@warning
       {Citation `\@citeb' on page \thepage \space undefined}}%
    {\setbox\z@\hbox{\global\@tempcntc0\csname b@\@citeb\endcsname\relax}%
     \ifnum\@tempcntc=\z@ \@citeo\@tempcntb\m@ne
       \@citea\def\@citea{,}\hbox{\csname b@\@citeb\endcsname}%
     \else
      \advance\@tempcntb\@ne
      \ifnum\@tempcntb=\@tempcntc
      \else\advance\@tempcntb\m@ne\@citeo
      \@tempcnta\@tempcntc\@tempcntb\@tempcntc\fi\fi}}\@citeo}{#1}}
\def\@citeo{\ifnum\@tempcnta>\@tempcntb\else\@citea\def\@citea{,}%
  \ifnum\@tempcnta=\@tempcntb\the\@tempcnta\else
   {\advance\@tempcnta\@ne\ifnum\@tempcnta=\@tempcntb \else \def\@citea{--}\fi
    \advance\@tempcnta\m@ne\the\@tempcnta\@citea\the\@tempcntb}\fi\fi}
\catcode`@=12
\begin{document}

%****************************************************************

% Titlepage

%****************************************************************

\setcounter{page}{0}
\thispagestyle{empty}
\begin{titlepage}

% BUTP Nr.
\vspace*{-1cm}
\hfill \parbox{3.5cm}{BUTP-2000/04 \\ MPI-PhT/2000-11\\
March 13, 2000\\
%hep-ph/9811518\\
}   
\vfill

% Title
\begin{center}
  {\large {\bf
Gluon Fragmentation into Glueballs and \\
Hybrid Mesons}
      \footnote{Work
      supported in part by the Schweizerischer Nationalfonds.}  }
\vfill
% Authors
{\bf
    Peter Minkowski } \\
    Institute for Theoretical Physics \\
    University of Bern \\
    CH - 3012 Bern, Switzerland
   \vspace*{0.3cm} \\  
   and \vspace*{0.3cm} \\
{\bf
    Wolfgang Ochs } \\
    Max Planck Institut f\"ur Physik \\
    Werner Heisenberg Institut \\
    D - 80805 Munich, Germany\\
%   \vspace*{0.5cm} \\  

\end{center}

\vfill

\begin{abstract}
The constituent nature of candidate gluonic mesons can be studied by
comparing their production in quark and gluon jets.
The production rate for such mesons depends on the colour confinement 
processes at the end of the perturbative evolution. Whereas we expect
enhanced production of hybrids in the fragmentation region
of a gluon jet, the rate for glueballs depends on the 
relative importance of colour triplet and colour octet neutralization.
These neutralization processes can be studied 
independently in events with large rapidity gaps.
If octet processes turn out important 
the recently suggested lightest $J^{PC}=0^{++}$
glueball with mass around 1000 MeV should become visible already in the
spectrum of the leading charged particle pairs.

\end{abstract}

%\vspace{-15cm}

%\rightline{MPI-PhT/97-46}
%\rightline{hep-ph/9707393} 
%\rightline{January 22, 2000}

\vfill
\end{titlepage}

%****************************************************************

%Wolfgang Ochs\thanks{E-mail address: wwo@mppmu.mpg.de}\\

\newpage

\section{Introduction}
A characteristic prediction of QCD in the non-perturbative sector concerns
the existence of mesons with gluons as valence constituents, either purely
gluonic mesons, ``glueballs'', or ``hybrids'', the bound states of 
quark, antiquark and gluon. 

The search for glueballs is an important issue since the early days of QCD
\cite{HFPM} but their very
existence has not been definitely established
until today. Whereas there is general agreement
that the lightest
glueball should have quantum numbers $J^{PC}=0^{++}$ its mass remains
controversial. Moreover, as $q\overline q$ mesons exist with the same
quantum numbers the identification is difficult, in particular, as
there could be mixing between the different states as well.

The important way to distinguish between the possibilities is the study
of production rates and decay branching ratios which are expected to be
sensitive to the constituent structure. Glueballs are expected
to be produced preferentially in a gluon rich environment such as
$J/\psi\to \gamma+X$, central production in hadron hadron collisions
by double Pomeron exchange  and $p\overline p$ annihilation
into mesons near threshold. On the other hand, glueball production should be
suppressed in $\gamma\gamma$ collisions (see \cite{close} for survey).
All these reactions have been explored with
increasing accuracy in the past. 

Here we want to discuss another pair of reactions where glueball 
or hybrid production
could be enhanced or suppressed, respectively. 
In the fragmentation region of the quark jet the usual $q\overline q$
hadrons which carry the
initial quark as valence quark are enhanced, gluonic mesons are expected to
be suppressed. In analogy, one might expect gluonic mesons 
to be enhanced in the fragmentation region of gluon jets and
 $q\overline q$ states to be suppressed.
Correspondingly, one may study
the relevant particle systems ($\pi\pi,\ K\overline K,\ 
4\pi\ldots)$ either at large momentum fraction (Feynman $x$), or
with some advantages, in the head
of the jet beyond a large rapidity gap (length $\Delta y$). 
Of advantage are the similar kinematic conditions for both kinds of jets
and often the possibility of a measurement in the same experiment.
The own identity of gluon jets produced in hard collisions has been
established in various details according to the expectations from
perturbative QCD calculations  (for review, see \cite{ko}).
There are plenty of gluon jets in the $e^+e^-\to q \overline q g$ 
final states and in hard $pp $ collisions which makes this type of studies
promising. 

Experimental studies along this line have not been carried out so far 
although there is some history
to the idea to look for gluonic mesons in  gluon jets. 
Peterson and Walsh \cite{pw} suggested a 
non-perturbative model with a leading isoscalar cluster in the
gluon jet which could be composed of 
ordinary \qqbar  mesons like $\eta,\eta',
\omega,\phi$ or glueballs. 
An enhanced production of
$\eta'$ in gluon jets
because of its intrinsic $gg$ component has been 
suggested by Fritzsch \cite{hf} in the explanation of the large decay rate 
of $B$ into $\eta'$.
More recently, Roy and Sridhar \cite{rs,rw} 
suggested the search for glueballs in gluon jets at large $x$
 and considered a model with the
$g\to gb $  fragmentation being similar to $q\to \pi$ fragmentation.

Whereas an earlier study at LEP \cite{l3} presented
evidence for an enhanced $\eta$ production in gluon jets at large $x$ 
the more recent studies of $\eta,\ \eta'$ mesons
 \cite{aleph,opal} did not indicate any enhanced production
of isoscalar mesons in gluon jets.
No study has been published
yet for glueball or hybrid candidates.

Whether the suggested systematics is realized in nature depends on the
 (nonperturbative) colour neutralization mechanism and could be
quite different for ordinary \qqbar mesons, hybrids and glueballs.
We distinguish the 
colour triplet and colour octet neutralization and suggest studies
which should reveal the relative importance of both mechanisms
independently of the existence of glueballs.
In any case, we expect the enhanced production of hybrids 
in gluon jets -- if they really exist (see also \cite{wo}).

\section{Status of glueballs and hybrids}

%The mass of the glueballs is not easily deduced from the \qqbar spectroscopy
%and in a first study 
%\cite{HFPM} different scenarios with low mass (around 1 GeV) and high mass
%(above 1.5 GeV) have been distinguished.

Lattice calculations in quenched approximation 
(without sea quarks) put the lightest
glueball with quantum numbers $J^{PC}=0^{++}$ near a mass of 1600 MeV
\cite{Morning,wein}
(for a review, see \cite{michael}). 
First results from unquenched calculations \cite{lattunq}
are not yet conclusive but may
indicate a decrease of the glueball mass with decreasing quark mass.

On the other hand, within the QCD sum rule
approach, a lighter gluonic component in the $0^{++}$ spectrum is required near
1000 MeV \cite{nar} (earlier results in \cite{prenar}). Also the early
calculations in the MIT bag model \cite{bag}
prefer the low mass glueball around 1000 MeV.

In a recent phenomenological analysis \cite{mo} it has been proposed
that the lightest  $0^{++}$ glueball should be identified
with the very broad state
\begin{equation}
gb(0^{++}):\qquad M\ =\ 1000\ \rm{MeV},\qquad \Gamma\ = \ 500\ -\ 1000
\ \rm{MeV}
\label{gb0}
\end{equation}
which corresponds to both states $f_0(400-1200)$ and $f_0(1370)$ listed in the
particle data table \cite{pdg}. Whereas in some experimental
 mass spectra there is a separate peak near 1370 MeV
the phase shift analyses of elastic and inelastic $\pi\pi$ scattering
do not indicate the  phase movement corresponding to  an ordinary
Breit Wigner resonance centered at this mass value. Rather -- after removal
of the narrow $f_0(980)$ -- the $\pi\pi$ 
elastic scattering phase shifts below 1500 MeV 
are slowly rising and  pass through  $90^{\degr}$  near 1000 MeV; there 
is no indication of a second  resonance -- neither in the elastic \cite{mp}
nor in the inelastic channels \cite{mo}. 

Recent data on central production of particle pairs by the WA102 and GAMS
collaborations \cite{wa102pairs,gamspairs}
show again peaks in the mass spectra near  $f_0(1370)$; the measured angular
distribution moments should now allow a judgement in favour
of the Breit-Wigner resonance
phase movement or against it from the interference with $f_2(1270)$;
the same applies also to
peripheral $\pi^0\pi^0$ production in $\pi p$ production by the E852
collaboration \cite{e852}. Relative phases may also be obtained in the
$4\pi$ final states measured by WA102 \cite{wa1024pi}. Such
analyses have the potential to further clarify whether
$f_0(1370)$ is a reasonably narrow ($\Gamma\sim 200-300$ MeV)
Breit-Wigner resonance or -- as we suggested \cite{mo} -- a component of a
yet broader object.

The  broad state $gb(1000)$ in (\ref{gb0}) is found 
consistent with most expectations for production and decay of a glueball
and is therefore a respectable candidate. The main decay mode is $\pi\pi$,
but also $K\overline K$ and $\eta\eta$ above the respective thresholds
and possibly $\pi\pi e^+e^-$ \cite{km}. The identification of $f_0(1500)$
as lightest glueball is not supported by the finding of opposite signs
for the decay amplitude into $K\overline K$ and $\eta\eta$ \cite{mo}. 
 
There is even less experimental information about $0^{-+}$  and $2^{++}$
glueballs. Possible candidates according to the analysis of
ref. \cite{mo} are $\eta(1440)$ and $f_J(1710)$ with $J=2$.
 
In the sector of hybrid mesons the spin-exotic state $1^{-+}$ is of
particular interest. Lattice results yield mass values around 1900 MeV
for this hybrid composed of light quarks and gluon in quenched approximation
and also in the first attempt including sea quarks
\cite{ukqcd,bernard,sesam,michael}.

There is experimental evidence for resonances with exotic $1^{-+}$ quantum numbers
in the $\eta\pi$ system at masses around 1400 and 1600 MeV (for a 
recent summary,
see Chung \cite{chung}). This result is lower than the lattice expectation by
around 500 MeV. This discrepency could be due to an incomplete calculation 
or the observed states are not hybrid but other exotic states, for example 
$q\overline q q\overline q$ states.

\section{Hadron production and colour neutralization}

An energetic  quark or gluon emerging from  a hard
collision process will generate a parton cascade
by subsequent gluon bremsstrahlung and quark pair
production.
The extension of such a cascade for a 100 GeV jet in space can be rather
large and exceed 100 f (see, for example \cite{dkmt,os}). The formation of
colour singlet systems should proceed during the evolution whenever
the separation of colour charges exceeds the confinement 
length $R_c\sim 1$f. Two types of neutralization processes are possible
(see Fig. 1).
\begin{figure*}[t]
\begin{center}
%lower left. upper right positions
%\mbox{\epsfig{file=ochs1.ps,width=7.0cm,bbllx=3.2cm,bblly=9.2cm,bburx=18.cm,bbu
\mbox{\epsfig{file=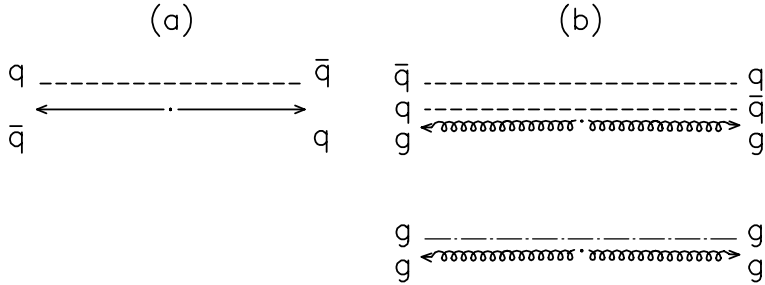,width=10cm}}
\end{center}
%\vspace{-0.7cm}
\caption[]{
The colour neutralization (a) of an initial $q\overline q$ pair 
by $\overline q q$ and (b) of an initial $gg$ pair by either  double 
colour triplet 
$q\overline q$ or by colour octet $gg$}
\label{fig1}
\end{figure*}

1. Colour triplet neutralization.\\ 
Consider the production of a massive
$ q\overline q$ pair in a colour singlet state 
in its restframe corresponding to 
separating colour triplet
charges  (possibly accompanied by secondary 
 partons from bremsstrahlung). 
The colour field between the primary quarks can be neutralized eventually by
the (nonperturbative)
 production of a soft quark antiquark pair.
 This process of triplet neutralization may repeat itself for the 
various branchings inside the quark gluon cascade  until
the total energy is carried by 
$q\overline q$ hadrons. 
Also possible is the formation of  hybrid mesons from the gluons at the end
of the perturbative cascade and the soft  $q\overline q$.
An example for a quantitative description 
of soft nonperturbative $q\overline q$
production is given by the chromoelectric flux tube model \cite{cnn}
which considers pair production in a static field. A systematic treatment
for relativistic particles connecting to perturbation theory has not yet been
achieved.

2. Colour octet neutralization.\\
In case of a primary colour singlet $gg$ pair the field between the 
separating colour octet charges can be neutralized at the confinement 
distance  either by the production of a gluon pair or by the subsequent
production of two quark pairs. 
Only the mechanism with  gluon pair production could yield
pure gluonic bound states at the end. 

We introduce the relative probability of
triplet and octet neutralization by a phenomenological parameter
\begin{equation}
R_3\ =\ P_3/P_8.    \label{ratio83}
\end{equation}
%from $P_3=4/3$ and $P_8=3$ as $R_3=4/13$ 
In the flux tube model of ref. \cite{cnn} the octet process was considered
negligible assuming a large effective gluon mass $>1 $ GeV in expectation of
a heavy glueball. We consider here the possibility of a light glueball and also
of a non-negligable soft $gg$ production rate.

%Estimate of $R_3=4/9$ in fluxtube model ...  (?)

It is not obvious to what extent the two types of neutralization mechanisms
are realized in a particular
process. Common collision processes at particle accelerators
are initiated by quarks either as
constituents of external hadrons or after production through electromagnetic
or weak gauge bosons. A successive triplet neutralization by quark pairs
is then always possible even in a gluon dominated final state.
The octet neutralization could be an overall rare process
enhanced for particular kinematic configurations.

Here we suggest such a configuration:
consider the production of a hard gluon 
which travels without gluon radiation for a while forming a jet with a
large rapidity gap empty of hadrons. The probability for such events
decreases exponentially with the rapidity gap according to the Sudakov
formfactor \cite{os}. 
In this case the hard isolated gluon builds up an
octet field to the remaining partons which is not distorted by multiple
gluon emissions and related colour neutralization processes of smaller
rapidity range, so the $gg$ 
octet mechanism may be enhanced and 
will become clearly visible if it exists. 

The experimental test can be carried out independently of the
existence of glueballs. In case of $gg$ neutralization the total charge of
the head particles beyond the gap should be $Q=0$;  
on the other hand, the charge
distribution in case of double triplet neutralization should have a
component with charges $Q=\pm 1$, a situation which is also met
for an equal mixture
of quark and antiquark jets with a gap.
Therefore, if the colour octet mechanism exists, the charge
distribution for the gluon gap events should be a mixture of both
components according to the probability $R_3$ in (\ref{ratio83}). 

In this test the gap $\Delta y$ should be large enough so that the leading
charges have approached a limiting distribution; only the charges which 
correspond to the combination of the leading parton and the soft
neutralizing parton (or partons) are present. In this limit multiple exchanges
through the gap and leading charges $|Q|\geq 2$ should be absent.   

If the enhanced neutral component from octet neutralization
is not observed we would not expect a preferred source of glueballs
in gluon jets either. In this case glueballs would be produced through
their mixing with quarkonium states in any collision process
and not preferentially through their valence glue component.

\section{Limiting charge distribution beyond the gap}

For illustration we give an estimate of the limiting charge distribution 
for sufficiently large $\Delta y$
in the head of the jet beyond the gap 
for the case of triplet neutralization. 
We assume that the charge is determined 
in the quark jet by the charge of
the leading quark and an average soft antiquark  and in case of  gluon jet by
the average charge of the soft quarks and antiquarks. 
In our estimate
we assume that soft $u,d,s$ quarks are produced with 
relative probabilities $\gamma,\gamma,\gamma_s$ 
and we take $\gamma=0.4,\gamma_s=1-2\gamma=0.2$
as in ref. \cite{ff} in the example (i.e. the production ratio $s/u=0.5$). 
More generally, $\gamma$ could depend on the energy scale of the separating
partons and we compare also with the extreme cases $\gamma=1/2\ (s/u=0)$
and $\gamma=1/3\ (s/u=1)$.

The charge distribution in the head of the gluon jet is then given by
\begin{equation}
p_g(Q)\ = \  P_3 p_3^g(Q) +P_8p_8^g(Q)
\label{Qgjet}
\end{equation}
where the charge distributions for triplet and octet neutralization
 $p_3^g(Q)$ and  $p_8^g(Q)$ are given
in Table 1. As seen from the table, the probabilities vary within 10\%
for the different assumptions on the strange quark ratio $s/u$.
From the measurement of the charge
distribution (\ref{Qgjet}) one can determine the ratio $R_3$ in
(\ref{ratio83}), given the parameter $\gamma$.

The corresponding analysis can be done as well in quark jets with a gap
to check the overall picture and study the parameter $\gamma$.
 The charge
distribution in the head of the jet for different quarks and the
quark-antiquark average relevant to $e^+e^-$ events is given in Table 2.
For a superposition of different quark (anti-quark)
jets with probabilities $\hat P_q$ 
the charge distribution is given by
\begin{equation}
\hat P_{<q>}(Q)\ = \ \sum_q \hat P_q \ p_3^q(Q). \label{quarkcharge}
\end{equation} 
There are some
differences between the gluon jet with triplet neutralization
and the $q\overline q$ average jet, but both are clearly separated from the
octet case which can therefore easily be identified with this method. 

\begin{table}[ht]
\caption{Leading charge distribution  $p_3^g(Q)$ and  $p_8^g(Q)$
in a gluon jet 
in case of triplet neutralization by soft $u,d,s$ quarks and anti-quarks
for different values of the $s/u\equiv \gamma_s/\gamma$ ratio
and octet neutralization by gluons.
 }
\[
  \begin{array}{cccccc}
\hline
  &  \mbox{triplet}\ p_3^g(Q) & && & \mbox{octet}\ p_8^g(Q) \\
\hline 
 & s/u=(1-2\gamma)/\gamma
 & s/u=0.5&  s/u=0 & s/u=1.0  & g
  \\
\hline 
p_c^g(Q=0) & 1-2\gamma+2\gamma^2 & 0.52 & 0.5 & 0.56  & 1 \\
p_c^g(Q=1) & \gamma (1-\gamma) &  0.24  & 0.25 & 0.22 & 0 \\
p_c^g(Q=-1) & \gamma (1-\gamma) & 0.24 & 0.25 & 0.22  & 0 \\
 \hline 
  \end{array} \]
 \label{tab1}
\end{table} 
\begin{table}[ht]
\caption{Leading charge distribution $p_3^q(Q)$ in quark jets;
numerical values for $s/u=0.5$.}
\[
  \begin{array}{cccccccccc}
\hline 
& u \mbox{ or } c & & (u+\overline u)/2 & d \mbox{ or } b& & (d+\overline d)/2
     & s & &(s+\overline s)/2
  \\
\hline 
P_3(Q=0) &\gamma   & 0.4 &  0.4 & \gamma  & 0.4 &  0.4 & 1-\gamma &  0.6 &
   0.6\\  
P_3(Q=1) &1-\gamma & 0.6 &  0.3 & 0       & 0.0 &  0.3 &   0      &  0.0 &
   0.2\\
P_3(Q=-1)& 0       & 0   &  0.3 & 1-\gamma& 0.6 &  0.3 & \gamma   &  0.4 &
   0.2  \\
 \hline 
  \end{array} \]
 \label{tab2}
\end{table} 

\section{Fragmentation Region of Gluon and Quark Jets}
 
The inclusive spectra of hadrons and other observables are well
described by the corresponding quantities at the parton level provided the
perturbative evolution is continued down to small values of the $k_T$ cutoff
in the cascade $Q_0\sim \Lambda_{QCD}$ (this phenomenological approach 
is called ``local
parton hadron duality'', for reviews, see \cite{dkmt,ko}). In this picture
the $x$-distribution of hadrons in the gluon jet is steeper than in the
quark jet because of the stronger semi-soft gluon radiation in the gluon
jet.

As to the fragmentation region of the quark jet 
it is by now well established that the 
hadrons which carry the primary quark as a valence quark dominate at large
momentum fraction $x$ as expected in the parton model
\cite{feynman}.  
The leading
hadron follows the distribution of the primary quark if it is
color-neutralized by a soft anti-quark. 

In the gluon jet, in case of triplet neutralization,  the leading particles 
are the hadrons formed by the primary gluon and the soft $q\overline q$ 
pairs. If there are hybrid mesons they may be formed in the fast
colour singlet $q\overline qg$ system, alternatively, the fast hadrons
are ordinary $q\overline q$ mesons formed at the end of the parton cascade
after having absorbed all gluon energy.
If the octet mechanism is at work
the leading gluon may also form a glueball.
On the other hand, in the quark jet neither the hybrid nor the glueball will
be leading if -- as we assume -- the leading quark is only neutralized in
colour by a soft antiquark. These properties are summarized in Table 3.

\begin{table}[ht]
\caption{Production of leading hadrons in the jet}
\[
  \begin{array}{ccccc}
\hline 
& \mbox{neutralization} &  \overline{q} q &  
  \mbox{hybrid} & \mbox{glueball} 
  \\
\hline 
 \mbox{quark jet} & \mbox{triplet} & \mbox{yes} &  \mbox{no}  &
       \mbox{no}\\
  \mbox{gluon jet} & \mbox{triplet} & \mbox{yes} &  \mbox{yes}  &
       \mbox{no}\\
                   & \mbox{octet} & \mbox{no} &  \mbox{no}  &  \mbox{yes}\\
  \hline 
  \end{array} \]
 \label{tab3}
\end{table} 

The constituent nature of a particular particle or resonance 
according to Table 3 can best 
be studied by comparing its production rate 
 in the fragmentation region
of quark and gluon jets. 
One possibility is the study of the respective mass spectra at large $x$ 
(see also \cite{rs}); there may be considerable non-resonant background. 
Another possibility is the study of particles 
beyond the rapidity gap. For large gaps when the limiting charge
distribution is reached the background should  be small.
As argued above the rapidity cut may enhance the possibility
for the octet mechanism. 

The comparison between quark and gluon jet is best carried out 
for similar kinematic configuration: the same length $\delta y$ for the head
of the jet beyond
the gap in a frame with the nearest jets in angle 
$\geq 90\dgr$.

For the gluon jet 
in $e^+e^-\to 3$ jets such a preferred frame
is reached after transforming first
into the restframe of $q\overline q$ and then
boosting along the $q\overline q$ direction until the $g$ jet is
perpendicular (see Fig. 2).
\begin{figure*}[ht]
\begin{center}
%lower left. upper right positions
%\mbox{\epsfig{file=ochs1.ps,width=7.0cm,bbllx=3.2cm,bblly=9.2cm,bburx=18.cm,bbu
\mbox{\epsfig{file=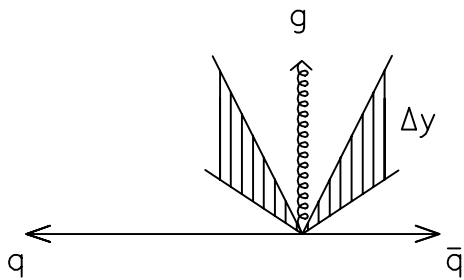,width=6cm}}
\end{center}
%\vspace{-0.7cm}
\caption[]{
Convenient frame to analyse  gluon jets with a rapidity
 gap $\Delta y$ in three jet events 
in $e^+e^-$ annihilation.}  
\label{fig2}
\end{figure*}
    
In order to 
correct for the different inclusive parton distributions in quark and
gluon jets and also in case of different available phase space
it is best to normalize the rates 
to a well known $q\overline q$ hadron $h_{norm}$ ($\rho,\ f_2\ldots$)
of comparable mass. Along this way 
we define the gluon factor $F_g$ for a hadron $h$
\begin{equation}
F_g(h/h_{norm})
    \ =\ \frac{\sigma(g\to h)/\sigma(g\to h_{norm})}
            {\sigma(q\to h)/\sigma(q\to h_{norm})}.
\label{gluon factor}
\end{equation}
Apparently, for a $q\overline q$ hadron 
the gluon factor $F_g\approx 1$ and for a gluonic
meson  $F_g \gg 1$.\footnote{
This quantity may be compared to the observable called  
``stickiness'' \protect\cite{chanowitz}, 
the ratio of radiative $J/\psi$ decay and $\gamma\gamma$ production for a
given hadron. In that case 
the very different phase space and barrier factors 
in both processes are taken into account in the definition 
 using an approximation appropriate for small masses; 
for high masses this approximation becomes unreliable and yields
rather large unrealistic ratios.}

According to our hypothesis about  $gb(1000)$ as being
the lightest binary glueball 
an effect $F_g > 1$ should be seen for the mass spectrum of $\pi^+\pi^-$ or
$\pi^0\pi^0$ around and below 1 GeV, i.e. in the simplest case already for
a neutral pair of charged particles. It may be compared with $\rho(770), \
f_2(1270)$. Other enigmatic states like the $f_0(980)$ can be tested for
their gluonic components as well. Heavier glueball candidates like
$f_0(1500)$ and $f_J(1710)$ may be studied through their $K\overline K$ decay
modes. Also the mass spectrum of all particles with total charge $Q=0$ 
in the head of the gluon jet
beyond the rapidity gap would be an interesting possibility for inclusive
glueball hunting.

\section{Conclusions}
The limiting charge distribution in the leading cluster 
of quark and gluon jets with sufficiently large
rapidity gap should reveal the relative importance of the octet
neutralization mechanism. The comparison of such events 
in quark and gluon jets could then provide some clues about the
constituent nature of candidate gluonic mesons 
with low background. 
Inclusive particle spectra did not prove any unusual
behaviour in both kinds of jets so far, but particle correlations could do.
Interesting tests are already possible with 
charged particle pairs.

\end{document}